\documentclass[letterpaper, 10 pt, conference]{ieeeconf}  % Comment this line out
\IEEEoverridecommandlockouts                              % This command is only needed if 
\usepackage{color}
\usepackage{graphics} % for pdf, bitmapped graphics files
\usepackage{epsfig} % for postscript graphics files
\usepackage{subfigure}
\usepackage{ulem}
\usepackage{times} % assumes new font selection scheme installed
\usepackage{amsmath} % assumes amsmath package installed
\usepackage{amssymb}  % assumes amsmath package installed
\usepackage{tikz}
\usepackage{pgfplots,amsmath}

\newcommand{\R}{\mathbb{R}}
\newtheorem{theorem}{Theorem}[section]

\newtheorem{definition}{Definition}[section]
\newtheorem{lemma}[definition]{Lemma}

\newtheorem{corollary}[definition]{Corollary}
\newtheorem{remarkth}[definition]{Remark}
\newenvironment{remark}{\begin{remarkth}\upshape}{\end{remarkth}}

\usepackage{amssymb}

\newcommand{\proa}{A^*G \mbox{$\;$}_{\tau^*} \kern-3pt\times_\alpha
G \mbox{$\;$}_\beta \kern-3pt\times_{\tau^*} A^*G}

\title{\LARGE \bf
Variational collision avoidance problems on Riemannian manifolds
}

\author{Mishal Assif, Ravi Banavar, Anthony Bloch, Margarida Camarinha and Leonardo Colombo \thanks{M. Assif (mishal\_assif@iitb.ac.in) is with Mechanical Engineering and R. N. Banavar (banavar@iitb.ac.in) is with the Systems and Control Enginerring, Indian Institute of Technology Bombay, Mumbai, Maharashtra 400076, India.  A. Bloch (abloch@umich.edu) is with Department of Mathematics, University of Michigan, 530 Church St. Ann Arbor, 48109, Michigan, USA.  M. Camarinha ( mmlsc@mat.uc.pt) is with CMUC, Department of Mathematics, University of Coimbra, 3001-501 Coimbra, Portugal . L. Colombo (leo.colombo@icmat.es) is with Instituto de Ciencias Matem\'aticas (ICMAT), Calle Nicol\'as Cabrera 13-15, 28049, Madrid, Spain.}}% {\tt\small msl.asf97@gmail.com, banavar@iitb.ac.in, leo.colombo@icmat.es, mmlsc@mat.uc.pt, abloch@umich.edu.}}%

\begin{document}

\maketitle
\thispagestyle{empty}
\pagestyle{empty}

%%%%%%%%%%%%%%%%%%%%%%%%%%%%%%%%%%%%%%%%%%%%%%%%%%%%%%%%%%%%%%%%%%%%%%%%%%%%%%%%
\begin{abstract}
In this article we introduce a variational approach to collision avoidance of multiple
agents evolving on a Riemannian manifold and derive necessary conditions for extremals.
The problem consists of finding non-intersecting trajectories of a given number of agents, among a set of admissible curves, to reach a specified
configuration, based on minimizing an energy functional that depends on the velocity, covariant acceleration and an artificial potential function used to prevent collision among the agents. The results are validated through numerical experiments on the manifolds $\mathbb{R}^{2}$ and $S^2$.
\end{abstract}

%%%%%%%%%%%%%%%%%%%%%%%%%%%%%%%%%%%%%%%%%%%%%%%%%%%%%%%%%%%%%%%%%%%%%%%%%%%%%%%%
\section{INTRODUCTION}

Path planning and collision avoidance of multiple agents have been areas of significant interest in the past few decades due to its broad applications in power networks, biological networks, social networks, mechanical networks and so on. Finding trajectories that take a set of agents from one configuration to another while avoiding collisions and minimizing some quantity like energy or time has been an important problem with applications in a variety of domains \cite{MMbook}. Distributed protocols were proposed for various agent networks, including general linear dynamical networks, nonlinear system networks, and mobile robotic networks in the last years \cite{Jorgebook}. Nevertheless there still exist some gaps in the literature in the bridge of knowledge between multi-agent systems and geometric mechanics. 

Calculus of variations in the large, as presented in Milnor \cite{Milnor}, has been exploited in the past for various applications. In Crouch and Silva Leite \cite{CroSil:91} the authors have used it to develop a theory of generalized cubic polynomials for dynamic interpolation problems on Riemannian manifolds. More recently, Bloch, Camarinha and Colombo \cite{BlCaCoCDC} have used these variational methods to solve obstacle avoidance problems on Riemannian manifolds. In this article, inspired by the recent work \cite{BlCaCoCDC}, we seek to extend this method to find necessary conditions for optimal trajectories of multiple agents on a Riemannian manifold that seek to achieve a specified configuration while avoiding collisions among themselves. Specifically, the problem studied in this paper consists of finding non-intersecting trajectories of a given number of agents, among a set of admissible curves, to reach a specified
configuration and minimizing an energy functional that depends on the velocity, covariant acceleration and an artificial potential function used to prevent collision. To solve the problem, we employ techniques from calculus of variations on Riemannian manifolds taking into account that the problem under study can be seen as a higher order variational problem \cite{blochcrouch2}, \cite{MargaridaThesis}, \cite{CMdD}, \cite{BlochHussein}.
 
The article is organized as follows. In section 2, we introduce some concepts from Riemannian geometry relevant to the rest of the article. In the next section, we define the variational collision avoidance problem on Riemannian manifolds and derive necessary conditions for the existence of extrema. In the following section we extend our analysis to Lie groups endowed with a left-invariant metric. We show how to apply the results of this work in particular examples: two agents moving on a Euclidean space, the sphere, and the collision avoidance of multiple rigid bodies on $SO(3)$.

%%%%%%%%%%%%%%%%%%%%%%%%%%%%%%%%%%%%%%%%%%%%%%%%%%%%%%%%%%%%%%%%%%%%%%%%%%%%%%%%
\section{Preliminaries}
\subsection{Preliminaries on Riemannian Geometry}

Let $M$ be a smooth $ (C^\infty) $ Riemannian manifold with Riemannian metric denoted by $ <\cdot,\cdot>\ : T_{x}M \ \times \ T_{x}M \rightarrow \mathbb{R}$ at each point $x \in M $, where $T_{x}M$ is the tangent space of $M$ at $x$.  The length of a tangent vector is determined by its norm,
$||v_x||=\langle v_x,v_x\rangle^{1/2}$ with $v_x\in T_xM$, for each point $x\in M$.

%Vector fields are used to calculate the directional derivative of a function defined on a manifold. There is another object in the realm of differential geometry which does the same
%for the much broader object, namely Tensors. This operator is called \textit{connection} (linear, covariant, or affine connection). 

A \textit{Riemannian affine connection} $\nabla$ on $M$, is a map that assigns to any two smooth vector fields $X$ and $Y$ on $M$ a new vector field, $\nabla_{X}Y$, called the \textit{covariant derivative of $Y$ with respect to $X$} satisfying $$\nabla_{fX}Y=f\nabla_XY,\hbox{ and } \nabla_X(fY)=X(f)Y+f\nabla_XY$$ for all vector fields $X,Y\in\mathfrak{X}(M)$ and $f\in\mathcal{C}^{\infty}(M)$, where $\mathfrak{X}(M)$ denotes the set of vector fields on $M$. For the properties of $\nabla$, we refer the reader to \cite{Boothby, bookBullo,Milnor}.

Consider a vector field $W$ along a curve $x$ on $M$. The $n$-th order covariant derivative of $W$ along $x$ is denoted by $\displaystyle{\frac{D^n W}{dt^n}}$ with $n\geq1$. We denote by $\displaystyle{\frac{D^{n+1} x}{dt^{n+1}}}$ the $n$-th order covariant derivative along $x$ of the velocity vector field of $x$, $n \geq 1$.

Given vector fields $X$, $Y$
and $Z$ on $M$, the vector field $R(X,Y)Z$ given by \begin{equation}\label{eq:CurvatureTensorDefinition}
R(X,Y)Z=\nabla_{X}\nabla_{Y}Z-\nabla_{Y}\nabla_{X}Z-\nabla_{[X,Y]}Z
\end{equation}  defines the \textit{curvature tensor} of $M$, where $[X,Y]$ denotes the \textit{Lie bracket} of the vector fields $X$ and $Y$. $R$ is trilinear in $X$, $Y$ and $Z$ and a tensor of type $(1,3)$. Hence for vector fields $X,Y,Z,W$ on $M$ the curvature tensor satisfies (\cite{Milnor}, p. 53)
\begin{equation}\label{curvformula}\langle R(X,Y)Z,W\rangle=\langle R(W,Z)Y,X\rangle.\end{equation}

%Let the Riemannian connection be denoted by $ \nabla \ : \mathfrak{X}(M) \times \mathfrak{X}(M) \rightarrow \mathfrak{X}(M)$ . 
Let $S$ be a submanifold of $M$ and $\Omega\subset M$ be the set of all $C^{1}$ piecewise smooth curves $x : [0,T] \ \rightarrow M$ such that $x(0)$, $\frac{dx}{dt}(0)$ and $x(T) \in S$ are fixed, and $\frac{dx}{dt}(T) \in T_{x(T)}S$. The set $\Omega$ is called the \textit{admissible set}. For the class of curves in $\Omega$ we introduce the $C^1$ piecewise smooth \textit{one parameter admissible variation} of a curve $x \in \Omega$ by $\alpha : (-\epsilon,\epsilon) \times [0,T] \rightarrow M ;(r,t) \mapsto\alpha(r,t) = \alpha_r(t)$ that satisfy $\alpha_{0} = x$ and $\alpha_r \in \Omega$, for each $r \in (-\epsilon, \epsilon) $. 
 
 The \textit{variational vector field} associated to an admissible variation $\alpha$ is a $C^1$-piecewise smooth vector  field $X$ along $x$ defined by $$X(t) = \frac{D}{\partial r}\Big{|}_{r=0}\alpha(r,t) \in T_{x(t)}\Omega$$ verifying the boundary conditions 
\begin{align*}
X(0)&=0, \quad X(T) = 0 \\
\frac{DX}{dt}(0)&=0, \quad \frac{DX}{dt}(T) \in T_{x(T)}M
\end{align*} where the tangent space of $\Omega$ at $x$ is the vector space $T_{x}\Omega$ of all $C^1$ piecewise smooth vector fields $X$ along $x$ verifying the former boundary conditions.

\begin{lemma}[\cite{Milnor}, p.$52$]\label{curvature_lemma}
The one parameter variation satisfies
\begin{equation*}
\frac{D}{\partial r}\frac{D^2\alpha}{\partial t^2} = \frac{D^2}{\partial t^2}\frac{\partial \alpha}{\partial r} + R \Big( \frac{\partial \alpha}{\partial r},\frac{\partial \alpha}{\partial t} \Big)\frac{\partial \alpha}{\partial t}.
\end{equation*}
\end{lemma}

Next, assume that $M$ is an $n$-dimensional complete Riemannian manifold. In this context the Riemannian distance between two points in $M$ can be defined by means of the Riemannian exponential on $M$, that is,
$$d(q,p)=\|\mbox{exp}_q^{-1}p\|.$$ We need to guarantee that the exponential map $\mbox{exp}_q$ is a local diffeomorphism, so  we assume that the point $p$ must belong to a convex open  ball around $q$.   If we consider the geodesic from $p$ to $q$ given by
$\gamma_{p,q}(s)= \mbox{exp}_p(s\, \mbox{exp}_p^{-1}q)$, $s\in[0,1]$, then, because $\displaystyle{\Big{\|}\frac{d \gamma_{p,q}}{d s}(s)\Big{\|}}$ is independent of $s$, we can write $$d^2(p,q)=\int_{0}^{1}\Big{\|}\frac{d \gamma_{p,q}}{d s}(s)\Big{\|}^2\, ds.$$

The proof of the following lemmas can be found in \cite{do1992riemannian}.

\begin{lemma}
Let $M$ be a complete smooth Riemannian manifold. Then $ d(p, q) = \| \mbox{exp}^{-1}_{p}q \|^2$ is well defined $\forall p, q \in M$.  \\
Moreover, if $\alpha : (-\epsilon, \epsilon) \rightarrow M$ is a smooth curve,
\begin{equation*}
\frac{\partial}{\partial r}d(p, \alpha(r))\Big{|}_{r=0} = -\Big{\langle} {\frac{\partial \alpha}{\partial r}}(0), \mbox{exp}^{-1}_{\alpha(0)}{p}\Big{\rangle}
\end{equation*}
\label{exp_map_lemma}
\end{lemma}
\begin{lemma}
If $X_i(t)$ is smooth vector field along $x_i(t) \in \Omega_i$, such that
\begin{equation*}
X_i(T) = 0 \hbox{ and } \frac{dX_i}{dt}(T) \in T_{x_i(T)}S
\end{equation*}
then $\alpha(r,t) = \hbox{exp}(rX_i(t))$ is an admissible variation of $x_i(t)$ whose variational vector field is $X_i$.
\label{var_vecfield_lemma}
\end{lemma}
\section{The variational collision avoidance problem on Riemannian manifolds}

\iffalse
\textcolor{red}{Is the graph theory introduced below essential since we do not make use of it later on ?}
%
Using graph theory, we can model the communication topology among $n$ agents in the formation.  A  graph $\mathcal{G}$ consists of a pair $(\mathcal{V},\mathcal{E})$, where $V=\{1,2,...,n\}$ is a finite nonempty set of nodes describing the kinematics of each single agent and $\mathcal{E}\subset \mathcal{V}\times\mathcal{V}$ denotes the set of edges of the graph, $p:=\frac{n(n-1)}{2}$ symmetric binary relations that link two agents. An edge $e_{(i,j)}$ denotes that node $j$ can obtain information from $i$. % All the neighbors of node $i$ are denoted as $\mathcal{N}_i:=\{e_{
%(j,|)}|(j,i)\in \mathcal{E}\}$, where we assume that $i\in\mathcal{N}_i$
\fi

\label{S:3}
Let $M$ be a complete smooth Riemannian manifold. Let $T$, $n$ and $k$ be positive real numbers. Consider $n$ agents evolving on $M$, and $(p_{0}^{i}, v_{0}^{i})$, with $i=1,2...,n$,  points in $TM$ corresponding to the initial positions and velocities of the agents. 

For each $i=1,\ldots,n$, consider the set $\Omega_{i}\subset M$ of all $C^1$ piecewise smooth curve on $M$, $x_{i} : [0,T] \rightarrow M$ verifying the boundary conditions
\begin{align*}
x_i(0) &= p_{0}^{i}, \quad \frac{dx_i}{dt}(0) = v_{0}^{i} \\ x_i(T) &= p_T^i \in S, \quad \frac{dx_i}{dt}(T) \in T_{x_i(T)}S
\end{align*}
(the $n$ agents reach a specified point on the submanifold $S$ with velocity tangent to $S$)
and define the functional $J$ on $\acute{\Omega} = \Omega_{1}\times$...$\times\Omega_{n}$ 
\begin{align*}
J(x_1, x_2, ..., x_n) =& \frac{1}{2}\sum_{i=1}^{n} \int_{0}^{T} \Bigg( \Big{\|} \frac{D^2x_i}{dt^2}(t) \Big{\|}^2 + k\Big{\|}\frac{dx_i}{dt}(t)\Big{\|}^2  \\ &+  \sum_{j=1, j \neq i}^{n} F( \| \hbox{exp}_{x_j(t)}^{-1}x_i(t)\|^2)  \Bigg) dt
\end{align*} where $\hbox{exp}_{x} : V_{0} \subset T_{x}M \rightarrow M$ is the geodesic exponential map, which is a smooth diffeomorphism of some open set around $0 \in T_{x}M$ onto an open set around $x$, $ F : \mathbb{R} \rightarrow \mathbb{R}^{*}$ is a smooth function from the reals to the extended reals such that $F(0) = +\infty$.
%\todo{who is p}
The functional is constructed as the sum of a combination of the velocity and covariant acceleration of the individual trajectories regulated by a parameter $k$ and a function that penalizes collisions between the agents. 

\textbf{Problem:} The variational collision avoidance problem involves minimizing the functional $J$ among $\acute{\Omega}$.

In order to minimize the functional $J$ among the set $\acute{\Omega}$ we want to find curves $x \in \acute{\Omega}$ such that $J(x) \leq J(\tilde{x})$, for all admissible curves $\tilde{x}$ in a $C^1$ neighborhood of $x$.

\begin{remark}
Note that the factor $\frac{1}{2}$ multiplying $\displaystyle{\sum_{j=1, j \neq i}^{n} F( \| \hbox{exp}_{x_j(t)}^{-1}x_i(t)\|^2)}$ is to not count twice the same potential function for two agents to avoid collision between them. %Other possibility might be to consider $\displaystyle{\sum_{j>i} F( \| \hbox{exp}_{x_j(t)}^{-1}x_i(t)\|^2)}$ as in \cite{CoDi} for $i=1,\ldots,n$. \hfill$\diamond$

\end{remark}

%\emph{Theorem 3.1 :}
%
\begin{theorem}\label{theorem1}
Let $x_i \in \Omega_i$. If $\alpha$ is an admissible variation of $x_i$ with variational vector field $X_i$, then 
\begin{align*}
0=&\frac{d}{dr}J(\alpha_r)\Big{|}_{r=0}= \int_{0}^{T} \Bigg( \Big  \langle X_i,\frac{D^4 x_i}{dt^4}- k\frac{D^2x_i}{dt^2} \\& \hspace{2.5cm}+ R \Big(\frac{D^2 x_i}{dt^2}, \frac{dx_i}{dt} \Big)\frac{dx_i}{dt}\\ &- \sum_{j=1 ,j\neq i}^{n} F'(\| \hbox{exp}_{x_i  (t)}^{-1} x_j(t)\|^2)\hbox{exp}^{-1}_{x_i(t)}x_j(t) \Big \rangle \Bigg) \ dt \\ &+ \sum_{i=1}^{l} \Big{[}\Big\langle\frac{DX_{i}}{dt},\frac{D^2x_{i}}{dt} \Big \rangle+\Big\langle X_i, k \frac{dx_i}{dt} - \frac{D^3x_i}{dt^3} \Big \rangle\Big{]}^{t_{i}^{-}}_{t_{i-1}^{+}}.
\end{align*}
\end{theorem}
%
%\emph{Proof :}
\textit{Proof:} If $\alpha$ is an admissible variation of $x_i \in \Omega_i$ with variational vector field $X_i$, then
\begin{equation*}
\begin{split}
& \frac{d}{dr}J(\alpha_r) = \int_0^T  \Bigg( \Big \langle \frac{D}{dr} \frac{D^2 \alpha}{dt^2}, \frac{D^2 \alpha}{dt^2}\Big \rangle + k\Big \langle\frac{D^2 \alpha}{\partial r \partial t}, \frac{\partial \alpha}{\partial t} \Big \rangle \\ &+  \sum\limits_{j=1, j \neq i}^{n} F'(\| \hbox{exp}^{-1}_{x_{j}(t)}\alpha(t)\|^2)\frac{\partial}{\partial r}\|\hbox{exp}^{-1}_{x_j(t)}\alpha_r(t)\|^2 \Bigg)dt.
\end{split}
\end{equation*}
By lemma \ref{exp_map_lemma}
\begin{equation*}
\begin{split}
\frac{\partial}{\partial r}\|\hbox{exp}^{-1}_{x_j(t)}\alpha_r(t)\|^2 = - \Big \langle \hbox{exp}^{-1}_{\alpha_r(t)}x_j(t), \frac{\partial}{\partial r} \alpha_r(t) \Big \rangle
%\Big \langle T_{\alpha(t)}exp^{-1}_{x_{j}(t)}.\frac{\partial \alpha}{\partial r}(t), exp^{-1}_{x_{j}(t)}\alpha(t) \Big \rangle_{x_{j}(t)} = \Big \langle  \frac{\partial \alpha}{\partial r}(t), T_{exp^{-1}_{x_{j}(t)}\alpha(t)}exp^{x_{j}(t)}.exp^{-1}_{x_{j}(t)}\alpha(t) \Big \rangle_\alpha(t)
\end{split}
\end{equation*}
%\newpage
By lemma \ref{curvature_lemma} and the previous equation \\
\begin{equation*}
\begin{split}
& \frac{d}{dr}J(\alpha_r) = \int_0^T \Bigg( \Big \langle \frac{D^2}{dt^2}\frac{\partial \alpha}{\partial r}, \frac{D^2 \alpha}{dt^2}\Big \rangle \\ & + \Big \langle R\Big(\frac{\partial \alpha}{\partial r}, \frac{\partial \alpha}{\partial t}\Big) \frac{\partial \alpha}{\partial t}, \frac{D^2 \alpha}{\partial t^2} \Big \rangle +  k\Big \langle\frac{D^2 \alpha}{\partial r \partial t}, \frac{\partial \alpha}{\partial t} \Big \rangle \\ &- \sum\limits_{j=1, j \neq i}^{n} F'(\| \hbox{exp}^{-1}_{x_{j}(t)}\alpha_r(t)\|^2)\Big \langle  \frac{\partial\alpha_r(t)}{\partial r}, \hbox{exp}^{-1}_{\alpha_r(t)}x_j(t) \Big \rangle \Bigg)dt.
\end{split}
\end{equation*}

Integrating the first term by parts twice and the third term once, and applying Lemma \ref{curvature_lemma} to the second term, we obtain that
\begin{equation*}
\begin{split}
& \frac{d}{dr}J(\alpha_r) = \int_0^T \Bigg( \Big \langle \frac{\partial \alpha}{\partial r}, \frac{D^4\alpha}{dt^4} + R \Big( \frac{D^2\alpha}{dt^2}, \frac{\partial \alpha}{\partial t} \Big) \frac{\partial \alpha}{\partial t} - k\frac{D^2\alpha}{dt^2} \\ &- \sum\limits_{j=1, j \neq i}^{n} F'(\| exp^{-1}_{x_{j}(t)}\alpha(t)\|^2)\Big \langle  \frac{\partial}{\partial r}\alpha_r(t), \hbox{exp}^{-1}_{\alpha_r(t)}x_j(t) \Big \rangle \Bigg) dt \\ &+ \sum_{i=1}^{l} \Big{[} \Big \langle \frac{D}{dt}\frac{\partial \alpha}{\partial r},\frac{D^2\alpha}{dt^2} \Big \rangle + \Big \langle \frac{\partial \alpha}{\partial r}, k \frac{\partial \alpha}{\partial t} - \frac{D^3\alpha}{\partial t^3} \Big \rangle\Big{]}^{t_{i}^{-}}_{t_{i-1}^{+}}
\end{split}
\end{equation*}
where the interval $[0,T]$ is partitioned as $ 0 = t_0 < t_1 < ...< t_l = T $ such that in each subinterval $x_i$ is smooth. \\
Taking $r = 0$ in the last equation, \\
%\begin{equation*}
%\begin{split}
\begin{align*}
& \frac{d}{dr}J(\alpha_r)\Big{|}{r=0} = \int_0^T \Bigg( \Big \langle X_i, \frac{D^4x_i}{dt^4} \\&\hspace{1.5cm}+ R \Big( \frac{D^2x_i}{dt^2}, \frac{\partial x_i}{\partial t} \Big) \frac{\partial x_i}{\partial t}  - k\frac{D^2x_i}{dt^2} \\ & - \Big( \sum\limits_{j=1, j \neq i}^{n} F'(\| \hbox{exp}^{-1}_{x_{j}(t)}x_i(t)\|^2). \hbox{exp}^{-1}_{x_i(t)}x_j(t) \Big)  \Big \rangle \Bigg) dt \\ & \sum_{i=1}^{l} \Big{[} \Big \langle \frac{DX_i}{dt},\frac{D^2x_i}{dt^2} \Big \rangle + \Big \langle X_i, k \frac{dx_i}{dt} - \frac{D^3x_i}{dt^3} \Big \rangle\Big{]}^{t_{i}^{-}}_{t_{i-1}^{+}}. 
\end{align*}$\hfill\square$
%\end{split}
%\end{equation*}
%
%
\begin{theorem}\label{mainTh}
If $\acute{x} \in \acute{\Omega}$ is a local minimizer of $J$, then $ \forall i \in {1,2...,n} $
\begin{enumerate} %[label=(\roman*)]
\item \begin{align*}&\frac{D^4 x_i}{dt^4} + R \Big(\frac{D^2 x_i}{dt^2}, \frac{dx_i}{dt} \Big)\frac{dx_i}{dt} - k\frac{D^2x_i}{dt^2}\\ &= \sum\limits_{j=1 ,j\neq i}^{n} F'(\| \hbox{exp}_{x_j (t)}^{-1} x_i(t)\|^2).(\hbox{exp}^{-1}_{x_i (t)} x_j(t))\end{align*}
\item $x_i$ is smooth on $[0,T]$
\item $\frac{D^2x_i}{dt^2}(T) \perp T_{x_{i}(T)}S$
\end{enumerate}
\end{theorem}
\proof
Assume $\acute x \in \acute\Omega$ is a local minimizer of $J$. Consider a variation of $\acute x$, $\acute\alpha_{r,i}(t) := (x_1(t),.. ,\alpha_{r,i}(t),... ,x_n(t))$, where $\alpha_{r,i}(t)$ is an admissible variation of $\Omega_i$ with variational vector field $X_i$. Then $\frac{d}{dr}J(\alpha_{r,i})\Big{|}_{r=0} = 0 \ \forall i \in 1,2,...n$. \par
Let us consider $X_i$ defined as
\begin{align*}
%\begin{split}
 & f \Big[ \frac{D^4x_i}{dt^4} + R\Big(\frac{D^2x}{dt^2},\frac{dx}{dt}\Big)\frac{dx}{dt} - k\frac{D^2x}{dt^2}
\\
 & - \Big( \sum\limits_{j=1 ,j\neq i}^{n} F'(\| \hbox{exp}_{x_i (t)}^{-1} x_j(t)\|^2).(\hbox{exp}^{-1}_{x_i (t)} x_j(t)) \Big)  \Big]
%\end{split}
\end{align*} where $f$ is a smooth real valued function on $[0,T]$ such that $f(t_i) = f'(t_i) = 0$ and $ f(t) > 0, t \neq t_i, i = i,...,l$.
So, we have \begin{align*}
0=&\frac{d}{dr}J(\alpha_r)\Big{|}_{r=0} = \int_0^T  \Bigg( f \| \frac{D^4x_i}{dt^4}\\& \hspace{3.2cm}+ R \Big( \frac{D^2x_i}{dt^2}, \frac{d x_i}{d t} \Big) \frac{dx_i}{dt}- k\frac{D^2x_i}{dt^2}
\\
&- \sum\limits_{j=1, j \neq i}^{n} F'(\| \hbox{exp}^{-1}_{x_{i}(t)}x_j(t)\|^2)\hbox{exp}^{-1}_{x_{i}(t)}x_j(t) \|^2  \Bigg) dt
\end{align*}
Since $f(t)$ is greater then zero outside a set of measure zero,
\begin{equation*}
\begin{split}
\Big\| \frac{D^4x_i}{dt^4} +& R \Big( \frac{D^2x_i}{dt^2}, \frac{d x_i}{d t} \Big) \frac{d x_i}{d t} - k\frac{D^2x_i}{dt^2} - \\  & \Big( \sum\limits_{j=1, j \neq i}^{n} F'(\| \hbox{exp}^{-1}_{x_{i}(t)}x_j(t)\|^2)\hbox{exp}^{-1}_{x_{i}(t)}x_j(t) \Big) \Big\| = 0
\end{split}
\end{equation*}
from which statement 1 follows. \par
Now, choose $X_i \in T_{x_i}\Omega_i$ such that \\
\begin{equation*}
\begin{split}
X_i(t_j) &= \frac{D^3x_i}{dt^3}(t_j^+) - \frac{D^3x_i}{dt^3}(t_j^-) \quad \forall j = 1,..,l-1 \\
\frac{DX_i}{dt}(t_j) &= \frac{D^2x_i}{dt^2}(t_j^-) - \frac{D^2x_i}{dt^2}(t_j^+) \quad \forall j = 1,..,l-1 \\
X_i(T) &= \frac{DX_i}{dt}(T) = 0
\end{split}
\end{equation*}
Therefore,
\begin{equation*}
\begin{split}
0=&\frac{d}{dr}J(\alpha_r)\Big{|}{r=0}=\sum_{i=1}^{l-1} \Big \| \frac{D^2x_i}{dt^2}(t_j^-) - \frac{D^2x_i}{dt^2}(t_j^+) \Big \|^2 \\ &+ \Big \| \frac{D^3x_i}{dt^3}(t_j^+) - \frac{D^3x_i}{dt^3}(t_j^-) \Big \|^2 =  0
\end{split}
\end{equation*}
which implies that
\begin{equation*}
\begin{split}
\frac{D^2x_i}{dt^2}(t_j^-) = \frac{D^2x_i}{dt^2}(t_j^+) \qquad \frac{D^3x_i}{dt^3}(t_j^+) = \frac{D^3x_i}{dt^3}(t_j^-)
\end{split}
\end{equation*}
Since $x_i$ is a $C^1$ curve with continuous covariant derivatives  up to order 3, $x_i$ is $C^3$ on $[0,T]$. But, we have shown that $x_i$ is the solution of a fourth order smooth ODE, which means the fourth derivative can be expressed as a smooth function of derivatives upto order 3. The $k^{th}$ order derivative can be expressed as a smooth function of derivatives upto order $k-1$, and so by induction, $x_i$ is smooth on $[0,T]$. Hence, statement 2 follows.

When $X_i(T) = 0, \ \frac{DX_i}{\partial t}(T) = \frac{dX_i}{dt}(T)$. Now, choose $X_i \in T_{x_i}\Omega_i$ such that
\begin{equation*}
\begin{split}
X_i(T) = 0, \hbox{ and } \frac{DX_i}{dt}(T) = \Pi_{T_{x_i(T)}S}\left(\frac{D^2x_i}{\partial t^2}(T)\right)
\end{split}
\end{equation*}
where $ \Pi_{T_{x_i(T)}S}V$ is the orthogonal projection onto $T_{x_i(T)}S$ of $V \in T_{x_i(T)}M$. Since
 $X_i(T) = 0,\ \frac{dX_i}{dt}(T) = \frac{DX_i}{\partial t}(T) \in T_{x_i(T)}S$. By lemma \ref{var_vecfield_lemma}, $X_i$ is the variational vector field of an admissible variation. Therefore,
\begin{equation*}
\begin{split}
\frac{d}{dr}J(\alpha_r)\big{|}{r=0} & = \Big \langle  \Pi_{T_{x_i(T)}S}\left(\frac{D^2x_i}{\partial t^2}(T)\right), \frac{D^2x_i}{\partial t^2}(T) \Big \rangle = 0 \\
&\implies \Pi_{T_{x_i(T)}S}\left(\frac{D^2x_i}{\partial t^2}(T)\right) = 0
\end{split}
\end{equation*}
Hence statement 3 holds. \hfill$\square$
\iffalse
If $\Gamma^{k}_{ij}(x) = 0 \ \forall x \in S$, then $\frac{DX_i}{\partial t}(T) = \frac{dX_i}{dt}(T)$. Choose $X_i \in T_{x_i}\Omega_i$ such that \\
\begin{equation*}
\begin{split}
X_i(T) = \Pi_{T_{x_i(T)}S}(\frac{D^3x_i}{dt^3} - k\frac{dx_i}{dt}) \qquad  \frac{DX_i}{dt}(T) = \Pi_{T_{x_i(T)}S}(\frac{D^2x_i}{\partial t^2}(T))
\end{split}
\end{equation*}
Therefore
\begin{equation*}
\begin{split}
\at{\frac{d}{dr}J(\alpha_r)}{r=0} & = \Big \langle  \Pi_{T_{x_i(T)}S}(\frac{D^3x_i}{dt^3} - k\frac{dx_i}{dt}), \frac{D^3x_i}{dt^3} - k\frac{dx_i}{dt} \Big \rangle = 0 \\
&\implies \Pi_{T_{x_i(T)}S}(\frac{D^3x_i}{dt^3} - k\frac{dx_i}{dt}) = 0
\end{split}
\end{equation*}
Hence, 3.2 (iv) follows.
\fi

\begin{remark}
We now make some remarks on the previous result for the following scenarios:
\begin{itemize}
\item If $M$ is an $n$-dimensional manifold and $S$ is an $m$-dimensional submanifold, with $m < n$, then the conditions $x(0)$ and $\frac{dx}{dt}(0)$ being fixed gives $2n$ independent boundary conditions, the condition $x(T)$ being fixed gives another $n$ independent boundary conditions, and $\frac{dx}{dt}(T) \in T_{x(T)}S$ gives yet another $n-m$ boundary conditions. Theorem \ref{mainTh} (3) gives an additional $m$ boundary conditions. The fourth order boundary value problem given by Theorem \ref{mainTh} (1) along with the $4n$ boundary conditions forms a well posed problem. 
\item If we were to change the endpoint boundary condition to be $\frac{d x_i}{dt}(T) = v_i$ (i.e $\frac{d x_i}{dt}(T)$ is a fixed vector), then it easily follows that conditions (1) and (2) of Theorem \ref{mainTh} still hold, but condition (3) does not hold. This still results in $4n$ boundary conditions. 
\item If we consider endpoint boundary conditions $x_i(T)\in S$ (i.e $x_i(T)$ is not fixed, but can be any element in $S$) and $\frac{dx_i}{dt}(T)\in T_{x_i(T)}S$, all the conditions in Theorem \ref{mainTh} still hold. But, in this case, we have $2n$ initial conditions, the endpoint conditions $x(T) \in S$ gives $n-m$ conditions, and $\frac{dx}{dt}(T) \in T_{x(T)}S$ gives $n-m$ boundary conditions. These $4n-2m$ conditions, along with the $m$ conditions given by Theorem \ref{mainTh} (3) gives only $4n-m$ boundary conditions. We expect to find another $m$ conditions, similar to Theorem \ref{mainTh} (3), in this case. We explore such an extension of Theorem \ref{mainTh} to this situation in a future work as we comment in Section \ref{lastsection}.
\end{itemize}
\end{remark}

\subsection{Example: Planar agents on an Euclidean space}\label{example1}
\begin{figure}
\centering
\includegraphics{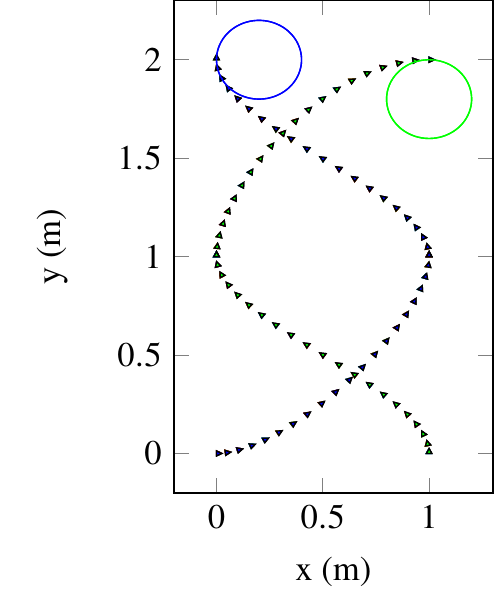}
\caption{Optimal Path traced by two agents evolving on $\R^2$.} \label{fig:M1}
\end{figure}

We consider the case of 2 agents evolving on $M = \mathbb{R}^2$ endowed with the Euclidean Riemannian metric. At time $T =0$, the first agent (agent blue in Figure \ref{fig:M1}) is at $p_0^1 = (0,0)$ with velocity $v_0^1 = (1,0)$, and the second agent (agent green in Figure \ref{fig:M1}) is at $p_0^2 = (1,0)$ with velocity $v_0^2 = (0,1)$. At time $T =1$ the first agent is at $p_1^1 = (1,1)$ with velocity $v_1^1 = (0,1)$, and the second agent is at $p_1^2 = (0,1)$ with velocity $v_1^2 = (0,1)$.  At time $T =2$ the first agent is at $p_2^1 = (0,2)$ with velocity $v_2^1 \in T_{p_2^1}S_1$, and the second agent is at $p_2^2 = (1,2)$ with velocity $v_2^2 \in T_{p_2^2}S_2$. Here, $S_1 = \{(z^1,z^2) \in \R^2 | \ \|(z^1,z^2)-(0.2, 2) \|_2 = 0.2\}$, and $S_2 = \{(z^1,z^2) \in \mathbb{R}^2 | \ \|(z^1,z^2)-(1, 1.8) \|_2 = 0.2\}$.  Note that in this particular example, $S$ (the submanifold
in Theorem \ref{mainTh}) is the disjoint union of the two circles shown in Figure \ref{fig:M1}. Here we have taken $F : (0,\infty) \rightarrow (0,\infty)$ given by  $\displaystyle{F(x) = \frac{1}{x}}$, and the value of $k=0$. Note that at time $T=1$, the boundary conditions are that mentioned in Remark 3.2. In this particular case, the argument of $F$ is the usual Euclidean distance between two points. We show in Figure \ref{fig:M1} some simulations of the trajectories given by Theorem \ref{mainTh}. Note that the trajectories exhibit the usual $S$-shape of cubic polynomials. \\
On $\mathbb{R}^2$ endowed with the Euclidean metric,  
\begin{align*}
exp^{-1}_{x_1}(x_2) = x_2 - x_1.
\end{align*}
We denote by $(x_i(t), y_i(t))$ the trajectory of the $i$-th agent. The conditions of Theorem \ref{mainTh} (1) translate to
\begin{align*}
x_1'''' &= \frac{x_1 - x_2}{((x_2-x_1)^2 + ((y_2-y_1)^2)^2}, \\
y_1'''' &= \frac{y_1 - y_2}{((x_2-x_1)^2 + ((y_2-y_1)^2)^2}, \\ 
x_2'''' &= \frac{x_2 - x_1}{((x_2-x_1)^2 + ((y_2-y_1)^2)^2}, \\ 
y_2'''' &= \frac{y_2 - y_1}{((x_2-x_1)^2 + ((y_2-y_1)^2)^2}. 
\end{align*}
Theorem \ref{mainTh} (3) gives
\begin{align*}
y_1''(2) = 0, \quad x_2''(2) = 0.
\end{align*}
along with the conditions
\begin{align*}
x_1(0) &= 0, \quad y_1(0) = 0,\quad x_2(0) = 1, \quad y_2(0) = 0, \\
x_1'(0) &= 1, \quad y_1'(0) = 0, \quad x_2'(0) = 0, \quad y_2'(0) = 1, \\
x_1(1) &= 1, \quad y_1(1) = 1, \quad x_2(1) = 0, \quad y_2(1) = 1, \\
x_1'(1) &= 0, \quad y_1'(0) = 1, \quad x_2'(1) = 0, \quad y_2'(0) = 1, \\
x_1(2) &= 0, \quad y_1(2) = 2, \quad x_2(2) = 1,\quad y_2(2) = 2. \\
x_1'(2) & = 0, \quad y_2'(2) = 0.     
\end{align*}

\subsection{Example: Agents on a 2-Sphere}

In this example, we consider 2 agents evolving on the $2$-sphere $S^2 = \{x \in \mathbb{R}^3 \mid \|x\|_2 = 1\} $ endowed with the induced Riemannian structure from $\R^3$.  We have taken fixed position and velocity conditions 
\begin{align*}
p_0^1 &= (0.7, 0.3, 0.648), \quad v_0^1 = (1, 0, -1.08), \\
p_1^1 &= (-0.7, 0.3, 0.648), \quad v_1^1 = (1, 0, 1.08), \\
p_0^2 &= (0.7, 0.6, 0.3873), \quad v_0^2 = (0, -1, 1.55), \\
p_1^2 &= (-0.7, 0.6, 0.3873), \quad v_1^2 = (0, 1, -1.55). 
\end{align*}
 As before, we have taken $F : (0,\infty) \rightarrow (0,\infty)$ given by  $\displaystyle{F(x) = \frac{1}{x}}$, and the value of $k=0$. Figure \ref{fig:M2} shows the trajectories satisfying Theorem \ref{mainTh} followed by the agents. Agent 1 and 2 follow the blue and brown trajectories respectively in \ref{fig:M2}. \\
For $x, y \in S^2$, 
\begin{align*}
exp^{-1}_x(y) &= \text{cos}^{-1}(\langle x, y \rangle) \frac{y - \langle x, y\rangle x}{\sqrt{1 - \langle x, y\rangle^2}}, \\
\|exp^{-1}_x(y)\| &= \text{cos}^{-1}(\langle x, y \rangle).
\end{align*}

For the purpose of computation, we parametrize $S^2$ with the $(\theta, \phi)$ coordinates as
\begin{align*}
x = (\text{sin}\theta \text{sin}\phi, \ \text{sin}\theta \text{cos}\phi, \ \text{cos}\theta)
\end{align*}
If $(\theta(t), \phi(t))$ is the coordinate representation of the curve $x(t)$, it can be shown that
\begin{align*}
\frac{D^4x}{dt^4} = \Big(&\theta'''' + (5\text{sin}2\theta) \theta'^2\phi'^2 +(1-7\text{cos}^2\theta)\theta''\phi'^2 \\& +(5-17\text{cos}^2\theta)\theta'\phi'\phi'' - (3\text{sin}\theta\text{cos}\theta)\phi''^2 \\& - (2\text{sin}2\theta)\phi'\phi''' + (\text{sin}\theta\text{cos}^3\theta)\phi'^4 \Big)\frac{\partial}{\partial \theta} + \\ & \Big(\phi'''' - 7\theta'\theta''\phi' -5\theta'^2\phi'' + (4\text{cot}\theta)\theta'''\phi' \\& + (6\text{cot}\theta)\theta''\phi'' + (4\text{cot}\theta)\phi'''\theta' \\& + (\text{sin}2\theta-\text{cot}\theta(5\text{cos}^2\theta-1))\theta'\phi'^3 \\ &- (6\text{cos}^2\theta)\phi'^2\phi'' - (2\text{cot}\theta)\phi'\theta'^3 \Big)\frac{\partial}{\partial \phi}, \\
 R \Big(\frac{D^2 x}{dt^2}&, \frac{dx}{dt} \Big)\frac{dx}{dt} = 0. 
\end{align*}

If $(\theta_i(t), \phi_i(t))$ denote the coordinate representation of the trajectory of the $i$-th agent, then \ref{mainTh} in local coordinates gives a 2 point boundary value problem in the variables $(\theta_i,\phi_i)$, which can be solved to obtain the optimal trajectories.
\iffalse
\begin{center}
\begin{figure}
\includegraphics[width=6cm, height=6cm]{sphere.eps}
\caption{Optimal Path traced by two agents evolving on $S^2$.} \label{fig:M2}
\end{figure}
\end{center}
\fi

\begin{figure}
\centering
\includegraphics{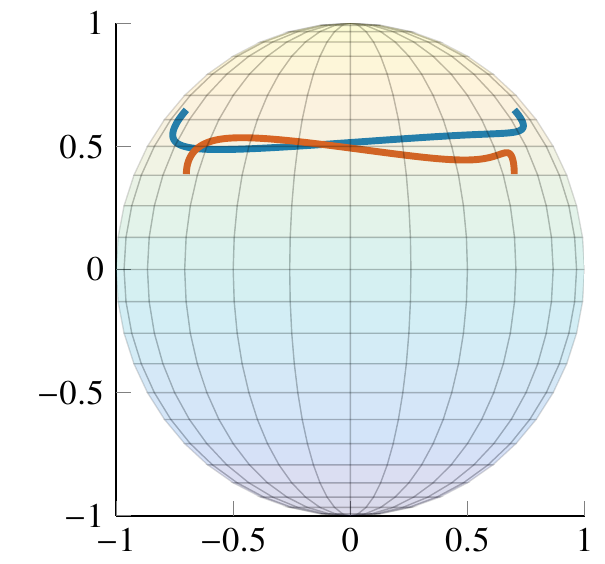}
\caption{Optimal Path traced by two agents evolving on $S^2$.} \label{fig:M2}
\end{figure}

%%%%%%%%%%%%%%%%%%%%%%%%%%%%%%%%%%%%%%%%%%%%%%%%%%%%%%%%%%%%%%%%%%%%%%%%%%%%%%%%

%\section{References}
%
%\bibliographystyle{model1-num-names}

\section{Variational collision avoidance problem on Lie Groups}
\label{S:4}

On a Lie group $G$, we can give an explicit notion of geodesic exponential.

Let $e$ be the identity element on $G$, and $\mathfrak{g}$ denotes the Lie algebra associated with $G$, that is, $\mathfrak{g}=T_{e}G$. A vector field $X\in\mathfrak{X}(G)$ is called \textit{left-invariant} if $T_{h}L_g(X(h))=X(L_g(h))=X(gh)$ for all $g,h\in G$. In particular for $h=e$ this means that a vector field $X$ is left-invariant if $\dot{g}=X(g)=T_{e}L_{g}\zeta$ for $\zeta=X(e)\in\mathfrak{g}$. Note that if $X$ is a left invariant vector field, then $\zeta=X(e)=T_{g}L_{g^{-1}}\dot{g}$.   

Given $\zeta \in \mathfrak{g}$, we denote by $X_\zeta$ the left-invariant vector field defined as $X_\zeta (g) = T_eL_g(\zeta), \forall g \in G$. Let $\phi^\zeta_t $ denote the flow of $X_\zeta$. Let $U$ be some neighborhood of $0 \in \mathfrak{g}$, the \textit{exponential function} $\hbox{exp}: U \subset \mathfrak{g} \rightarrow G$ is defined as $\hbox{exp}(\zeta) = \phi^\zeta_1(e)$. 

If $G$ is equipped with a Riemannian structure, it is not in general true that the 2 notions of an exponential coincide. Geodesics on a Lie group are not generally the flow of a left-invariant vector field on $G$. A connection on $G$ for which geodesics are flows of left-invariant vector fields is called a Cartan connection.

Any inner product on $\mathfrak{g}$ induces a left invariant Riemannian metric on $G$ (see \cite{bookBullo} p. 271). The restriction of such a Riemannian metric to $\mathfrak{g}$ will be denoted by $\widetilde{\nabla} : \mathfrak{g} \times \mathfrak{g} \rightarrow  \mathfrak{g}$. A Riemannian metric which is both left and right invariant is called bi-invariant. Unfortunately, the Riemannian metric being left-invariant does not guarantee the Levi-Civita connection it induces will be a Cartan connection. It can be shown that the Levi-Civita connection induced by a bi-invariant Riemannian metric on $G$ is also a Cartan connection (see \cite{sternberg2013curvature} p. 156). In fact, on a connected Lie group, the converse of the statement is also true, i.e. if the Levi-Civita connection induced by a left-invariant Riemannian metric is also a Cartan connection, then the Riemannian metric is bi-invariant. In the following discussion, we exclusively deal with the case where $G$ is endowed with a left-invariant Riemannian metric such that the Levi-civita connection it induces is also a Cartan connection.

Therefore, assume that $G$ endowed with a left-invariant  Riemannian metric $\langle\cdot,\cdot\rangle$, with $\mathbb{I}:\mathfrak{g}\times \mathfrak{g}\to\R$ the corresponding inner product on the Lie algebra $\mathfrak{g}$, a positive-definite symmetric bilinear form in $\mathfrak{g}$.  The inner product $\mathbb{I}$ defines  the metric $\langle\cdot,\cdot\rangle$ completely via left translation (see for instance \cite{bookBullo} pp. 273).

%The Levi-Civita connection $\nabla$ induced by $\langle\cdot,\cdot\rangle$ is an affine left-invariant connection and it is completely determined by its restriction to $\mathfrak{g}$ via left-translations. This restriction, denoted by $\stackrel{\mathfrak{g}}{\nabla}:\mathfrak{g}\times\mathfrak{g}\to\mathfrak{g}$,  is given by (see \cite{bookBullo} p. 271)

%\begin{equation}\label{restrictedconnection}\stackrel{\mathfrak{g}}{\nabla}_wu= \frac 12 [w,u]-\frac 12 \mathbb{I}^{\sharp}\left(\hbox{ad}_w^* \mathbb{I}^{\flat}(u)+\hbox{ad}_u^* \mathbb{I}^{\flat}(w)\right),\end{equation} where \hbox{ad}$^{*}:\mathfrak{g}\times\mathfrak{g}^{*}\to\mathfrak{g}^{*}$ is the co-adjoint representation of $\mathfrak{g}$ on $\mathfrak{g}^{*}$ and where $\mathbb{I}^{\sharp}:\mathfrak{g}^{*}\to\mathfrak{g}$, $\mathbb{I}^{\flat}:\mathfrak{g}\to\mathfrak{g}^{*}$ are the associated isomorphisms with the inner product $\mathbb{I}$ (see \cite{Boothby} for instance).

Let $x:I\subset\mathbb{R}\to G$ be a smooth curve on $G$. The body velocity of $x$ is the curve $v:I\subset\mathbb{R}\to\mathfrak{g}$ defined by $\displaystyle{v(t)=T_{x(t)}L_{x(t)^{-1}}\left(\frac{dx}{dt}(t)\right)}$.

Let $\{e_1,\ldots,e_n\}$ be a basis of $\mathfrak{g}$. We denote by $u_{L}$ the left-invariant vector field associated with $u\in\mathfrak{g}$. The body velocity of $x$ on the given basis is described  by $\displaystyle{v=\sum_{i=1}^n v_i e_i}$, where $v_1,\ldots, v_n$ are  the so-called pseudo-velocities
of the curve $x$ with respect to the given basis. The velocity vector can be written in terms of the pseudo-velocities as follows
\begin{equation}\label{admissibility}\frac{dx}{dt}(t)=T_eL_{x(t)}v(t)=\sum_{i=1}^n v_i(t) (e_i)_L(x(t)).\end{equation}

When the body velocity is interpreted as a control on the Lie algebra, equations \eqref{admissibility} give rise to the so called left-invariant control systems discussed in \cite{CoDi}. Therefore our analysis also includes this class of kinematic control systems.

To write the equations determining necessary conditions for existence of extrema in the variational collision avoidance problem, we must use the following formulas (see \cite{Altafini}, Section $7$ for more details).
% Let $x : I \subset \mathbb{R} \rightarrow G$ be a smooth curve on $G$. The body velocity of $x$ is the curve $v : I \subset R \rightarrow \mathfrak{g} $ defined by $v(t) = T_{x(t)}L_{x(t)^{-1}}\Big(\frac{dx}{dt}(t))\Big) $. Let ${e_1,...,e_n}$ be a basis of $\mathfrak{g}$. Consider the body velocity $v$ on the given basis, defined as $v = \sum\limits_{i=1}^{n}v_ie_i$. It follows that $\frac{dx}{dt}(t) = T_eL_{x(t)}v(t) = \sum\limits_{i=1}^{n}v_i(t) T_{e}L_{x_i(t)}(e_i) $. On such a lie group, the following formulas hold
\begin{align*}
\widetilde{\nabla}_vv &= \sum\limits_{i,j=1}^{n}v_iv_j\widetilde{\nabla}_{e_j}e_i,\quad \hbox{exp}^{-1}_{x}y = T_eL_x(exp_e^{-1}(x^{-1}y)),
 \\
\frac{D^2x}{dt^2} &= T_eL_x\Big(v' + \widetilde{\nabla}_vv\Big), \\
\frac{D^3x}{dt^3} &= T_eL_x\Big(v'' + \widetilde{\nabla}_{v'}v + 2\widetilde{\nabla}_{v}v' +\widetilde{\nabla}_v\widetilde{\nabla}_vv\Big), \\
\frac{D^4x}{dt^4} &= T_eL_x\Big(v''' + \widetilde{\nabla}_{v''}v + 3\widetilde{\nabla}_{v'}v' + 3\widetilde{\nabla}_{v}v'' + \widetilde{\nabla}_{v'}\widetilde{\nabla}_{v}v
\\
&+ 2\widetilde{\nabla}_{v}\widetilde{\nabla}_{v'}v + 3\widetilde{\nabla}_{v}v' + \widetilde{\nabla}_{v}v\Big), \\
&R\Big(\frac{D^2x}{dt^2} , \frac{dx}{dt} \Big)\frac{dx}{dt} = T_eL_x\Big(\widetilde{R}(v',v)v + \widetilde{R}(\widetilde{\nabla}_{v}v,v)v\Big), 
%&\textcolor{blue}{T_{exp^{-1}_{y}(x)}exp_y(exp^{-1}_{y}(x)) = T_eL_x\Big(exp^{-1}_{e}(y^{-1}x)\Big) }
\end{align*} where $\widetilde{R}$ denotes the curvature tensor associated with $\widetilde{\nabla}$. Thus, as a consequence of Theorem \ref{mainTh} (1) we have the following result

\begin{corollary}\label{corollary2}
 The equations giving necessary conditions for the existence of minimizers in the variational collision avoidance problem where agents are defined on a Lie group $G$ are
\begin{equation*}
\begin{split}
0=&v_i''' + \widetilde{\nabla}_{v_i''}v_i + 3\widetilde{\nabla}_{v_i'}v_i' + 3\widetilde{\nabla}_{v_i}v_i'' + \widetilde{\nabla}_{v_i'}\widetilde{\nabla}_{v_i}v_i \\ & + 2\widetilde{\nabla}_{v_i}\widetilde{\nabla}_{v_i'}v_i + 3\widetilde{\nabla}_{v_i}v_i' + \widetilde{\nabla}_{v_i}v_i + \widetilde{R}(v_i',v_i)v_i\\ & + \widetilde{R}(\widetilde{\nabla}_{v_i}v_i,v_i)v_i - k\Big(v_i' + \widetilde{\nabla}_{v_i}v_i\Big) \\ & - \Big( \sum\limits_{j=1,i\neq j}^{n} F'(\|\hbox{exp}^{-1}_{e}(x_{j}^{-1}x_{i})\|^2)\hbox{exp}^{-1}_{e}(x_{i}^{-1}x_{j}) \Big).
\end{split}
\end{equation*}
\end{corollary}
\subsection{Example: Collision avoidance of rigid bodies on $SO(3)$}
We clarify the notion of collision avoidance on $SO(3)$ as follows: The idea is to ensure that $n$ agents, 
each evolving on the manifold $SO(3)$, do not attain the same orientation at any given instant of time. 
In this particular example, 
we consider a collision avoidance problem for the motion of three rigid bodies in the space where the configuration space of each agent is the Lie group $G=SO(3)$.  

 Denote by $t\to R_i(t)\in SO(3)$ a curve for the $i^{th}$ agent, $i=1,2,3$ and let $I$ be the $(3\times 3)$-identity matrix. The columns of the matrix $R_i(t)$ represent the directions of the principal axis of the $i^{th}$ body at time $t$ with respect to some reference system. Let $\mathfrak{so}(3)=T_{I}SO(3)$ be the Lie algebra of the Lie group $SO(3)$, that is, the set of $3\times 3$ skew-symmetric matrices, \begin{align*}\mathfrak{so}(3)&=\{\dot{R}(0)|R(t)\in SO(3), R(0)=I\}\\&=\{\hat{\Omega}\in\mathbb{R}^{3\times 3}|\hat{\Omega} \hbox{ is skew-symmetric}\}.\end{align*} It is well know that (see \cite{Bl} for instance)  $\mathfrak{so}(3)\simeq\mathbb{R}^3$ using the isomorphism $$\hat{\Omega}_i(t)=\left(
  \begin{array}{ccc}
    0& -\Omega^3_i(t) & \Omega^2_i(t) \\
    \Omega^3_i(t) & 0 & -\Omega^1_i(t) \\
    -\Omega^2_i(t) & \Omega^1_i(t) & 0 \\
  \end{array}
\right)\simeq(\Omega_i^{1},\Omega_{i}^{2},\Omega_i^{3})$$ where $\Omega_i=(\Omega_i^{1},\Omega_{i}^{2},\Omega_i^{3})\in\mathbb{R}^{3}$.%\in\mathbb{R}^{3}.$$

We consider the basis of $\mathfrak{so}(3)$ represented by the canonical basis of $\mathbb{R}^3$ denoted by $\{e_1,e_2,e_3\}$ and  endow $SO(3)$ with the left-invariant metric  defined by the inner product $\displaystyle{\mathbb{I}=\sum_{k=1}^3J_k^{i}e^k\otimes e^k}$ where $J_k^i$ are the elements of the diagonal matrix defining the kinematics structure of the rigid body, the inertia moments, with $k=1,2,3$ for each $i=1,2,3$ and $\{e^{1},e^2,e^3\}$ the dual basis of $\{e_1,e_2,e_3\}$.

The Levi-Civita connection $\nabla$ induced by $\langle\cdot,\cdot\rangle$ determined by the inner product is completely determined by its restriction to $\mathfrak{so}(3)$ and given by (see for instance \cite{bookBullo} p. 281)
$$\widetilde{\nabla}_{v} z=\frac{1}{2}v\times z+\frac{1}{2}\left(\mathbb{I}^{-1}(v\times(\mathbb{I}z)+w\times(\mathbb{I}v)\right)$$ where $v=(v_1,v_2,v_3)\in\mathbb{R}^{3}$ and $z=(z_1,z_2,z_3) \in \mathbb{R}^3$.

For simplicity in the exposition we consider the case of a symmetric rigid body with $J_k^i=1$ with $i,k=\{1,2,3\}$. Then the formula above for the Levi-Civita connection reduces to
$\displaystyle{\widetilde{\nabla}_{v} z=\frac{1}{2}(v\times z).}$
By using equation \eqref{curvformula}, the  restriction of the curvature tensor  to  $\mathfrak{so}(3)$ is defined by
$\displaystyle{\widetilde{R}(v,z)w=-\frac 14 (v\times z)\times w}$
where $v,z,w \in \mathbb{R}^3$.

For the collision avoidance problem we consider $F:\mathbb{R}^{+}\to\mathbb{R}^{+}$ as $F(x)=\frac{1}{x}$ where the argument of $F$ is the usual Euclidean distance between two elements on $\mathfrak{so}(3)\simeq\mathbb{R}^{3}$, that is, $F(\|\mbox{exp}_{Q}^{-1}R\|^2)=\frac{1}{\|\mbox{exp}_{Q}^{-1}R\|^2}$ with $R,Q\in SO(3)$ and where $\exp:\mathfrak{so}(3)\to SO(3)$ representing the matrix exponential on $SO(3)$.

%Given that for each $i,j=1,2,3$, $$T_IL_{R}\left(\frac{1}{\|\mbox{exp}_{R_i}^{-1}R_j\|^2}\right)=\frac{\exp_{R_j}^{-1}R_i}{\|\mbox{exp}_{R_i}^{-1}R_j\|^4},$$
By Corollary \ref{corollary2} the necessary conditions for the normal extremals for the variational collision avoidance problem are determined by the solutions of the equation

\begin{align*}
v_i'''= &v_i''\times v_i + k v_i'+ \frac{3}{4}(v_i' \times v_i)\times v_i + \frac{3}{2}v_i'\times v_i \\ &+ \sum\limits_{j=1,i\neq j}^{3} \frac{\hbox{exp}^{-1}(R_j^{-1}R_i)}{\|\mbox{exp}^{-1}(R_i^{-1}R_j)\|^4}. %\\&+\frac{2\hbox{exp}_I^{-1}(R_3^{-1}R_1)}{\|\mbox{exp}_I^{-1}({R_1}^{-1}R_3)\|^4}+\frac{2\hbox{exp}_I^{-1}(R_2^{-1}R_3)}{\|\mbox{exp}_I(R_3^{-1}R_2)\|^4}.
\end{align*}
together with the equation $\dot{R}_i=R_iv_i$, and the boundary conditions $R_{0}=R_0^i$, $R_i(T)=R_T^i$, $v_i(0)=v_0^i$ and $v_i(T)=v^i_T$ for $i=1,2,3$.

%%%%%%%%%%%%%%%%%%%%%%%%%%%%%%%%%%%%%%%%%%%%%%%%%%%%%%%%%%%%%%%%%%%%%%%%%%%%%%%%
%\iffalse
\section{Conclusions and Future work}\label{lastsection}
We discussed the problem of collision avoidance of multi-agent systems on a complete Riemannian manifold and derived, from the point of view of calculus of variations, necessary conditions for the existence of extrema in the problem. We have shown how the main result can be applied for the particular case of a compact Lie group, a Euclidean space, and a non Lie group example, i.e., the $2$-sphere, $S^{2}$.

The study of these necessary conditions on symmetric spaces and reduction theories for variational problems has attracted considerable interest and has been carried out systematically by several authors. In future work we intend to extend the main results presented in this paper to this setting and explore numerical results for three dimensional agents as well as  explore extensions of Theorem \ref{mainTh} as we commented in Remark 3.2.
\section{Acknowledgments}
The research of A. Bloch was supported by NSF grant DMS-1613819 and AFOSR grant 17RT0038. The research of M. Camarinha was partially supported by the Centre for Mathematics of the University of Coimbra -- UID/MAT/00324/2013, funded by the Portuguese Government through FCT/MEC and co-funded by the European Regional Development Fund through the Partnership Agreement PT2020. The work of L. Colombo was partially supported by Ministerio de Economia, Industria y Competitividad (MINEICO, Spain) under grant MTM2016-76702-P and ``Severo Ochoa Programme for Centres of Excellence'' in R$\&$D (SEV-2015-0554) and Juan de la Cierva Incorporaci\'on Fellowship.

\end{document}